\documentstyle[prl,twocolumn,epsf,floats,aps]{revtex}
\begin{document}
\draft

\twocolumn[\hsize\textwidth\columnwidth\hsize\csname @twocolumnfalse\endcsname

\title{High-energy magnon dispersion and multi-magnon continuum in the 
two-dimensional Heisenberg antiferromagnet}

\author{Anders W. Sandvik$^1$ and Rajiv R. P. Singh$^{2}$}

\address{$^{1}$Department of Physics, University of Illinois at 
Urbana-Champaign, 1110 West Green Street, Urbana, Illinois 61801 \\
$^{2}$Department of Physics, University of California, Davis, 
California 95616}

\date{\today}

\maketitle

\begin{abstract}
We use quantum Monte Carlo simulations and numerical analytic continuation 
to study high-energy spin excitations in the two-dimensional $S=1/2$ 
Heisenberg antiferromagnet at low temperature. We present results 
for both the transverse $(x)$ and longitudinal $(z)$ dynamic spin structure 
factor $S_{x,z}({\bf q},\omega)$ at ${\bf q}=(\pi,0)$ and $(\pi/2,\pi/2)$. 
Linear spin-wave theory predicts no dispersion on the line connecting these
momenta. Our calculations show that in fact the magnon energy at $(\pi,0)$ 
is 10\% lower than at $(\pi/2,\pi/2)$. We also discuss the transverse and 
longitudinal multi-magnon continua and their relevance to neutron scattering 
experiments.
\end{abstract}

\pacs{PACS: 75.40.Gb, 75.40.Mg, 75.10.Jm, 75.30.Ds}
\vskip2mm]

It is now well established that the spin-wave theory of the two-dimensional 
Heisenberg model \cite{anderson} correctly describes the low-energy dynamics 
of layered antiferromagnets such as La$_{\rm 2}$CuO$_{\rm 4}$ 
\cite{neutrons1,chn}. Several methods have been used to calculate the quantum 
renormalization factor $Z$ of the spin wave velocity, $c=Zc_0$, with the 
result $Z \approx 1.18$ for $S=1/2$\cite{oguchi,singh,spinwave2,wiese,sandvik}.
Early neutron scattering experiments were consistent with this renormalization
over the entire Brilloin zone \cite{neutrons2}. However, recently more 
accurate measurements have shown clear deviations at high energies in 
La$_{\rm 2}$CuO$_{\rm 4}$ \cite{neutrons3} and other materials 
\cite{neutrons4}. In particular, spin-wave theory predicts no dispersion on 
the magnetic zone boundary, i.e., along the line of momenta ${\bf q}=
(\pi-x,x)$, whereas 
the experimental data show a significant variation in the excitation energy. 
This could be caused by interactions other than the nearest-neighbor 
super-exchange $J$ normally used to model the materials \cite{neutrons3,morr}.
The deviations could also be at least partially due to a failure of low-order
spin-wave theory to account for the high-energy dynamics of the Heisenberg 
model. A series expansion calculation \cite{singh2} indeed indicates 
a momentum dependent renormalization factor, with the magnon energy at 
${\bf q}=(\pi,0)$ approximately $7$\% lower than at $(\pi/2,\pi/2)$. This, 
however, is opposite to the trend found experimentally for 
La$_{\rm 2}$CuO$_{\rm 4}$ \cite{neutrons3}.

Another important issue, that has not yet been confronted experimentally, 
is the multi-magnon continuum expected to be present in the dynamic
structure factor $S({\bf q},\omega)$ above the single-magnon mode. The 
neutron scattering cross-section has been analyzed assuming only a 
delta-function representing the spin waves with dispersion $\omega_{\bf q}$
\cite{neutrons1,neutrons2,neutrons3,neutrons4}. 
The resolution of present experiments is not sufficient for detecting the 
presence of additional spectral weight, much less to determine its 
distribution. Theoretical calculations have so far also not been 
successful in accurately determining the full dynamic structure factor.
In order to provide guidance for improved fitting procedures it is imperative
to obtain quantitative estimates of the multi-magnon spectral features, in 
particular considering that experiments may soon become sufficiently accurate 
for detecting the continuum. Within spin-wave theory, correctly accounting 
for the interactions that give rise to the continuum is extremely complicated 
and  has led to contradictory results \cite{spinwave2,igarashi,canali}. 
Numerical calculations of $S({\bf q},\omega)$ are challenging because large 
lattices have to be used to converge to the thermodynamic limit, and the 
extraction of real-time dynamics from quantum Monte Carlo (QMC) data is 
difficult. Such calculations have therefore so far only given limited insights 
\cite{makivic,syljuasen}. The series work \cite{singh2} suggests a significant 
continuum but gives no information on its shape.

Here we address the issues of high-energy magnon dispersion and multi-magnon 
continuum using QMC calculations in a way that explicitly separates the 
transverse and longitudinal components of the dynamic structure factor. This 
allows us to more accurately determine both the magnon energy and the 
continuum part of the spectrum. We consider the Heisenberg model on a square 
lattice, defined in standard notation by the Hamiltonian
\begin{equation}
H = J \sum\limits_{\langle i,j\rangle} {\bf S}_i \cdot {\bf S}_j .
\label{ham}
\end{equation}
The coupling $J \approx 1500$ K in typical planar cuprates. On an infinite 
lattice the spin-rotational symmetry of this model is spontaneously broken at 
$T=0$ and long-range order develops along an axis that we take as the 
$z$-direction. The two-point correlations then have different longitudinal 
($z$) and transverse $(x,y)$ components. In a finite system the symmetry 
is not broken and all correlation functions are equal to the same 
rotational average. We want to access separately the transverse dynamic 
structure factor, which contains the single-magnon mode as well as a 
continuum, and the longitudinal component comprising only a multi-magnon 
continuum. To this end, we explicitly break the symmetry by applying
a staggered magnetic field; 
\begin{equation}
H \to H - h\sum_{i} (-1)^{x_i+y_i} S^z_i.
\end{equation}
We adjust the field strength $h$ so that the induced staggered 
magnetization equals the known value in the thermodynamic limit;
$m = |\langle S^z_i \rangle | = 0.307$ 
\cite{oitmaa,reger,singh,spinwave2,wiese,sandvik}. 
In the limit of large system sizes $h \to 0$ (at $T=0$) and the correlation 
functions become equal to their values in a system with spontaneously broken 
symmetry. 

In real layered cuprates there is a small 
inter-layer coupling and some degree of anisotropy, resulting in a 
finite-$T$ transition to an ordered state, in La$_{\rm 2}$CuO$_{\rm 4}$ at 
$T_{\rm N} \approx 300$ K. We here consider a low temperature, $\beta=J/T=32$, 
corresponding to approximately $50$ K, where the staggered magnetization in 
real materials is very close to the saturated $T=0$ value. Our staggered 
field $h$ can then also be viewed as a mean-field treatment of the 
inter-layer coupling \cite{scalapino}. 

Working with
lattices with $N= L\times L$ spins and periodic boundary conditions, we 
have used the stochastic series expansion method \cite{sse} to calculate 
the imaginary-time dependent spin-spin correlation function
\begin{equation}
G_\alpha ({\bf q},\tau) = 
\langle S^\alpha (-{\bf q},\tau) S^\alpha ({\bf q},0) \rangle ,
\end{equation}
where $\alpha = x,z$ and
\begin{equation}
S^a({\bf q},\tau) = {1\over \sqrt{N}} 
\sum_i {\rm e}^{-{\bf q} \cdot {\bf r}_i} 
{\rm e}^{-\tau H} S^\alpha_i {\rm e}^{\tau H} .
\end{equation}
The correlation function is related to the dynamic structure factor 
according to
\begin{equation}
G_\alpha ({\bf q},\tau) = {1\over \pi}\int_{-\infty}^\infty 
d\omega S_\alpha({\bf q},\omega){\rm e}^{-\tau \omega},
\label{gtos}
\end{equation}
which in principle can be inverted to yield the real-frequency dynamics from
the correlation function computed in the simulations. With simulation data 
affected by statistical fluctuations there are well known difficulties in 
carrying out this analytic continuation in practice, and one can only expect 
limited frequency resolution using, e.g., the maximum entropy method 
\cite{maxent}. Here we will instead assume reasonable functional forms for 
the transverse ($x$) and longitudinal ($z$) dynamic structure factors, with 
parameters that are adjusted to satisfy the equality (\ref{gtos}). With the 
QMC method used, derivatives of $G_\alpha ({\bf q},\tau)$ can also be directly
calculated \cite{ramanpaper} and impose additional constraints on
$S_\alpha({\bf q},\omega)$;
from Eq.~(\ref{gtos})
\begin{equation}
G^{(n)}_\alpha ({\bf q},\tau) = {(-1)^n\over \pi} \int_{-\infty}^\infty 
d\omega S_\alpha({\bf q},\omega)\omega^n {\rm e}^{-\tau \omega}.
\label{gtder}
\end{equation}
We use the first $(n=1)$ and $(n=2)$ second derivatives, which at $\tau =0$
correspond to the first two frequency moments of the spectrum. We also use 
the sum rule
\begin{equation} 
\chi_\alpha ({\bf q})= {2\over \pi}
\int_{-\infty}^\infty {d\omega \over \omega}S_\alpha ({\bf q},\omega),
\end{equation}
where $\chi_\alpha ({\bf q})$ is the static susceptibility
\begin{equation}
\chi_\alpha ({\bf q}) = \int_0^\beta d\tau G_\alpha ({\bf q},\tau).
\end{equation}
The spectrum also obeys the bosonic relation $S_\alpha ({\bf q},-\omega) = 
{\rm e}^{-\beta\omega}S_\alpha ({\bf q},\omega)$.

For the model forms of the transverse and longitudinal structure factors we 
take simple functions that reflect the expected gross spectral features.
The transverse component should include a delta-function at an energy
$\omega_{\bf q}$, representing the magnon, and a continuum that does not 
extend below $\omega_{\bf q}$ (temperature broadening of the high-energy
magnons should be insignificant at $\beta=32$). The longitudinal 
component is entirely in the continuum and can also not extend below 
$\omega_{\bf q}$. Both continua must decay to zero rapidly as 
$\omega \to \infty$, to ensure convergence of all frequency moments. 
We use
\begin{eqnarray}
S_x({\bf q},\omega) & = & A_1({\bf q}) \delta (\omega - \omega_q) + 
A_2({\bf q}) f_x({\bf q},\omega) \label{sxform}, \\
S_z({\bf q},\omega) & = & B({\bf q}) f_z({\bf q},\omega), \label{szform}
\end{eqnarray}
where the continua $f_{x,z}({\bf q},\omega)$ are given by
\begin{eqnarray}
f_x({\bf q},\omega) 
& = & r_x {\rm e}^{-(\omega -\nu)^2/2\sigma^2}, 
~~ (0~{\rm for}~ \omega < \omega_q) , \label{tcont} \\           
f_z({\bf q},\omega) 
& = & r_z (\omega-\omega_q)^p {\rm e}^{-a(\omega-\omega_q)^b},
~~ (0~{\rm for}~ \omega < \omega_q) \label{lcont},
\end{eqnarray}
where $r_x$ and $r_z$ are factors normalizing the frequency integrals
to unity. For clarity, we have suppressed the dependence of the parameters 
$\nu,\sigma,p,a$, and $b$ on ${\bf q}$. The static structure factors 
(the integrated spectral weights) are given by
\begin{eqnarray}
S_x({\bf q}) & = & \langle S^x(-{\bf q})S^x({\bf q})\rangle = 
A_1({\bf q}) + A_2({\bf q}), \\
S_z({\bf q}) & = & \langle S^z(-{\bf q})S^z({\bf q})\rangle = 
B({\bf q}).
\end{eqnarray}
In scattering experiments with unpolarized neutrons, a rotational average of 
the transverse and longitudinal dynamic structure factors is measured. 
The rotationally averaged static structure factor $S({\bf q})=2S_x({\bf q})+
S_z({\bf q})= 2A_1+2A_2+B$.

The values we find for the ``self-consistent'' staggered field (demanding 
$m=0.3070$ to an accuracy of better than $0.00005$) are $h/J=0.10652$ for
$L=4$, $0.022575$ for $L=8$, $0.005450$ for $L=16$, and $0.001615$ for $L=32$. 
With the QMC algorithm \cite{sse} implemented in the $z$-basis, and with the 
field also in the $z$-direction, the longitudinal correlations can be easily 
calculated. In order to compute the transverse correlations it is more 
efficient to apply the field in the $x$-direction, so that the measurements 
can be carried out with diagonal operators \cite{henelius}. Hence, we perform
two independent simulations for each lattice size.

\begin{figure}
\centering
\epsfxsize=7.2cm
\leavevmode
\epsffile{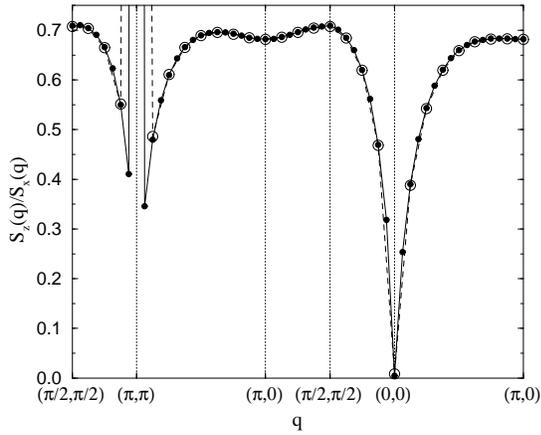}
\vskip1mm
\caption{Ratio of the longitudinal and transverse static structure factors
(integrated spectral weights) along a path in the Brilloin zone. Open
circles are for $L=16$ and solid ones for $L=32$.} 
\label{fig1}
\end{figure}

Fig.~\ref{fig1} shows our results for the ratio of the longitudinal and 
transverse spectral weights versus the momentum, along a standard path in 
the Brilloin zone. At ${\bf q}=(\pi,\pi)$ the ratio diverges (in the 
thermodynamic limit), reflecting the divergence of $S_z(\pi,\pi)$ due to
the long-range order. For ${\bf q} \to (0,0)$ and ${\bf q} \to (\pi,\pi)$ 
it approaches zero and hence the magnon completely exhausts the total 
spectral weight in these limits (other treatments have shown that the magnon 
exhausts the low-energy transverse spectral weight \cite{stringari}). The 
ratio is the largest at ${\bf q}=(\pi/2,\pi/2)$ where it is above $0.7$ ---
it is between $0.6$ and $0.7$ over much of the Brilloin zone. Hence, in
unpolarized neutron scattering experiments, $30-35$\% of the cross-section 
at the zone boundary is due to the longitudinal continuum.

Next we study the full dynamic structure factor, focusing on the zone-boundary
momenta ${\bf q} = (\pi,0)$ and $(\pi/2,\pi/2)$. The analytic continuation 
using fits to the functions (\ref{sxform}) and (\ref{szform}) was carried out 
using QMC imaginary-time data with a spacing $\Delta\tau=0.5$ up to $\tau = 4$
(the data become too noisy for higher $\tau$). For $L=4$, we have 
compared with exact diagonalization results and find that the magnon energy 
is very accurately reproduced (better than 0.5\%) but the weight in the magnon
is underestimated by about $5\%$ because the form of the continuum, 
Eq.~(\ref{tcont}), is not appropriate for a very small system where only 
a small number of significant delta-functions represent the continuum. 
Our results for larger systems should have smaller systematic errors. 

\begin{figure}
\centering
\epsfxsize=7.7cm
\leavevmode
\epsffile{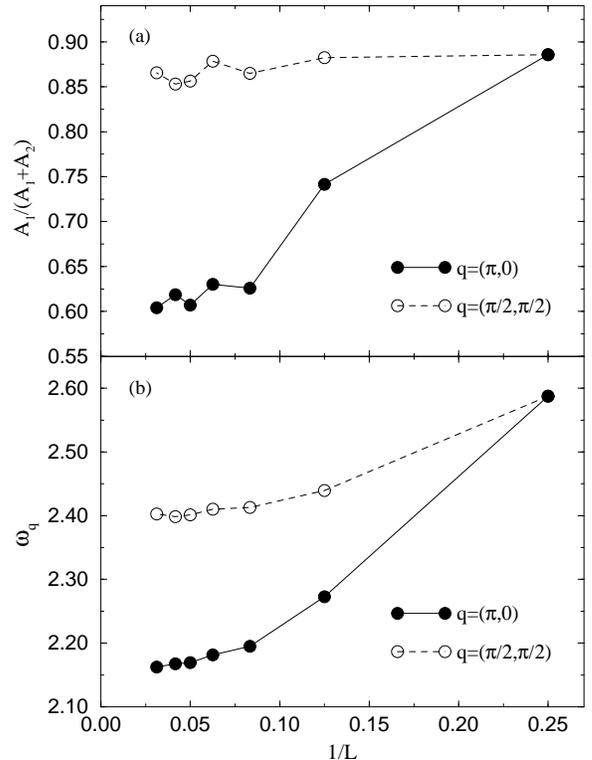}
\vskip1mm
\caption{Properties of the single-magnon part of the transverse dynamic
structure factor vs inverse system size ($L=4,8,12,16,20,24$, and $32$). 
(a) single-magnon fraction of the weight, (b) magnon energy.}
\label{fig2}
\end{figure}

Fig.~\ref{fig2} shows the size dependence of the magnon weight and
energy (the scatter of the data gives an indication of the statistical 
errors). The $L=4$ data is from exact diagonalization; in this case the 
two momenta are degenerate due to equivalence of this lattice
and a $2^4$ hyper-cube. Rough extrapolations in $1/L$ give the infinite-size
energy $\omega_{\bf q}/J \approx 2.15$ for ${\bf q}=(\pi,0)$ and $2.39$ for 
$(\pi/2,\pi/2)$. The relative weight of the magnon in $S_x({\bf q},\omega)$
is $\approx 60$\% at $(\pi,0)$ and $85$\% at $(\pi/2,\pi/2)$. Within
linear spin-wave theory, including the quantum-renormalization $Z=1.18$,
the energy $\omega_q/J = 2.36$ for both momenta. Hence, the results show that 
the interactions neglected in spin-wave theory have strong effects at 
${\bf q}=(\pi,0)$, lowereing the energy by almost $10$\% and transferring 
$\approx$ 40\% of the magnon weight into the continuum. At $(\pi/2,\pi/2)$ 
the excitation energy is instead slightly increased by the interactions and 
a much smaller fraction of the weight is in the continuum.

\begin{figure}
\centering
\epsfxsize=7.8cm
\leavevmode
\epsffile{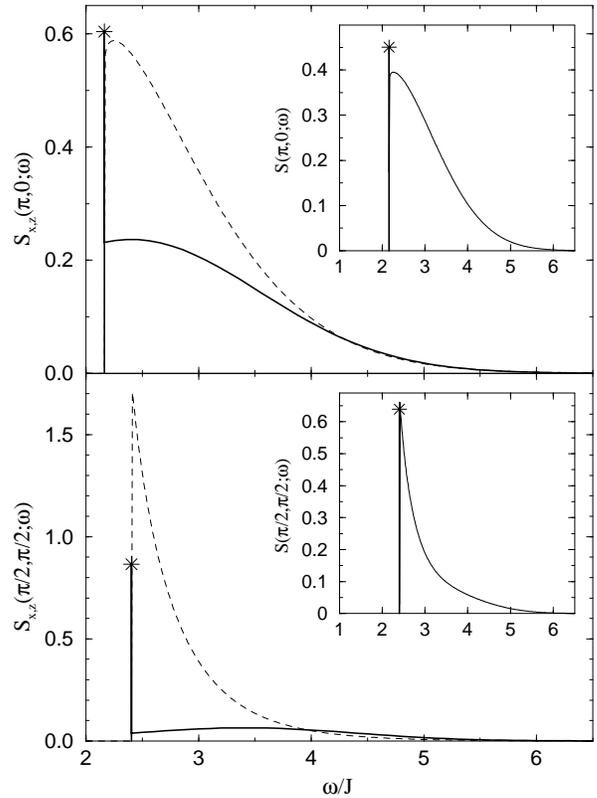}
\vskip1mm
\caption{The transverse (solid curves) and longitudinal (dashed curves)
dynamic structure factor at ${\bf q}=(\pi,0)$ (upper panel)
and $(\pi/2,\pi/2)$ (lower panel). $S_x({\bf q},\omega)$ and 
$S_z({\bf q},\omega)$ are normalized to $1$ and $S_z({\bf q})/S_x({\bf q})$ 
[see Fig.~\protect{\ref{fig1}}], respectively. The delta-function piece of 
$S_x({\bf q},\omega)$ is represented by a vertical line with height 
(indicated by a star) equal to its relative weight $A_1({\bf q})$. The 
insets show the rotationally averaged spectra, $S({\bf q},\omega)= 
2S_x({\bf q},\omega) + S_z({\bf q},\omega)$.}
\label{fig3}
\end{figure}

Fig.~\ref{fig3} shows the magnon as well as the transverse and longitudinal 
continua for $L=32$. The rotationally averaged structure factor measured in 
neutron scattering experiments with unpolarized neutrons is also shown. The 
exponent $p$ in the longitudinal continuum, Eq.~(\ref{lcont}), is small and 
positive, $p \alt 0.05$, giving a spectral weight concentrated close to the 
magnon. The longitudinal continuum is significantly narrower at $(\pi/2,\pi/2)$
than at $(\pi,0)$. For both momenta, the transverse continuum is broad and 
centered further above the magnon. All these spectral features are consistently
present for system sizes $L \ge 8$. As seen in the insets of Fig.~\ref{fig3}, 
for $L=32$ only $\approx 45$\% of the rotationally averaged spectral weight 
is in the magnon mode at $(\pi,0)$, increasing to $\approx 65$\% at 
$(\pi/2,\pi/2)$. The magnon weight decreases only slightly for even 
larger systems, as can be inferred from Fig.~\ref{fig2}.

In previous work using QMC and analytic continuation \cite{makivic,syljuasen},
the magnon energy was extracted as the first moment of the rotationally
averaged relaxation function, $F({\bf q},\omega) \sim S({\bf q},\omega)/
\omega$. Using this procedure, we find the moment $2.53$ at $(\pi,0)$ and 
$2.58$ at $(\pi/2,\pi/2)$, i.e., both significantly above the actual magnon 
energies (the first moment of $S({\bf q},\omega)$ is higher yet; approximately
$2.66$ for both momenta). This is clearly due to the large spectral weight 
in the continuum. There is potentially a similar problem in the analysis of 
neutron scattering data. Since the present-day energy resolution is such that 
a non-negligible part of the continuum that we have found here would be
included in an experimental fit to a single resolution-broadened peak, the 
extracted magnon energy may be too high. This effect would be 
particularly important at and close to ${\bf q}=(\pi,0)$, where our results 
show that less than $50$\% of the total weight is in the magnon peak and the
continuum is relatively broad. 

Experimental results do appear to show the presence of some weight consistent 
with a continuum above the resolutioned broadened magnon peak \cite{neutrons3},
but the statistics is not sufficient for extracting its size and shape. 
Our calculations should be useful for analyzing future experiments with higher
resolution. Such experiments will be very important for determining the 
significance of other interactions in the antiferromagnetic cuprates beyond 
the nearest-neighbor exchange $J$. The significant effects of magnon 
interactions that we have found at ${\bf q}=(\pi,0)$ may also have important
consequences for the broadening of the Raman spectrum \cite{canali} and the 
momentum space anisotropy of the photoemission spectrum \cite{wells}.

\vskip1mm
We would like to thank G. Aeppli, L. Balents,  G. Sawatzky, and D. Scalapino 
for stimulating discussions. This research was supported by the NSF under 
grants No.~DMR-97-12765 and DMR-99-86948. The work was started at the 
Institute of Theoretical Physics at UC Santa Barbara, under support of NSF
grant No.~PHY-94-07194. Most of the numerical calculations were carried out 
on the SGI Origin2000 system the NCSA.
\null

\vskip-4mm
\begin{references}
\null\vskip-20mm\null
\bibitem{anderson} 
P. W. Anderson, Phys. Rev. {\bf 86}, 694 (1952).

\bibitem{neutrons1}
D. Vaknin {\it et al.}, Phys. Rev. Lett. {\bf 58}, 2802 (1987);
G. Shirane {\it et al.}, {\it ibid.} {\bf 59}, 1613 (1987);
G. Aeppli {\it et al.},  {\it ibid.} {\bf 62}, 2052 (1989).

\bibitem{chn} 
S. Chakravarty, B. I. Halperin, and D. R. Nelson, Phys. Rev. Lett. {\bf 60},
1057 (1988); Phys. Rev. B {\bf 39}, 2344 (1989).

\bibitem{oguchi} 
T. Oguchi, Phys. Rev. {\bf 117}, 117 (1960).

\bibitem{singh}
R. R. P. Singh, Phys. Rev. B {\bf 39}, 9760 (1989).

\bibitem{spinwave2}
C. J. Hamer, Z. Weihong, and P. Arndt, Phys. Rev. B {\bf 46}, 6276 (1992);
J. Igarashi, Phys. Rev. B {\bf 46}, 10763 (1992);
C. M. Canali and M. Wallin, Phys. Rev. B {\bf 48}, 3264 (1993).

\bibitem{wiese}
U.-J. Wiese and H.-P. Ying, Z. Phys. B {\bf 93}, 147 (1994);
B. B. Beard and U.-J. Wiese, Phys. Rev. Lett. {\bf 77}, 5130 (1996).

\bibitem{sandvik}
A. W. Sandvik, Phys. Rev. B {\bf 56}, 11678 (1997).

\bibitem{neutrons2}
S. Hayden {\it et al.}, Phys. Rev. Lett. {\bf 67}, 3622 (1991).

\bibitem{neutrons3}
R. Coldea {\it et al.} (unpublished).

\bibitem{neutrons4}
Y. J. Kim et al Phys. Rev. Lett. {\bf 83}, 852 (1999); 
H. M. R{\o}nnow {\it et al.} (unpublished). 

\bibitem{morr}
D. K. Morr, Phys. Rev. B {\bf 58}, R587 (1998).

\bibitem{singh2}
R. R. P.Singh and M. P. Gelfand, Phys. Rev. B {\bf 52}, 15695 (1995).

\bibitem{igarashi}
J. Igarashi and A. Watabe, Phys. Rev. B {\bf 43}, 13456 (1991).

\bibitem{canali}
C. M. Canali and S. M. Girvin, Phys. Rev. B {\bf 45}, 7127 (1992).

\bibitem{makivic}
M. S. Makivic and M. Jarrell, Phys. Rev. Lett. {\bf 68}, 1770 (1992).

\bibitem{syljuasen}
O. F. Sylju{\aa}sen and H. M. R{\o}nnow, cond-mat/0003350.

\bibitem{oitmaa}
J. Oitmaa and D. D. Betts, Can. J. Phys. {\bf 56}, 897 (1978).

\bibitem{reger}
J. D. Reger and A. P. Young, Phys. Rev. B {\bf 37}, 5978 (1988).

\bibitem{scalapino}
D. J. Scalapino, Y. Imry, and P. Pincus, Phys. Rev. B {\bf 11},
2042 (1975).

\bibitem{sse}
A. W. Sandvik, Phys. Rev. B {\bf 59}, R14157 (1999).

\bibitem{maxent} 
M. Jarrell and J. E. Gubernatis, Phys. Rep. {\bf 269}, 133 (1996).

\bibitem{ramanpaper} 
A. W. Sandvik, S. Capponi, D. Poilblanc, and E. Dagotto,
Phys. Rev. B {\bf 57}, 8478 (1998).

\bibitem{henelius}
P. Henelius, A. W. Sandvik, and S. M. Girvin, 
Phys. Rev. B {\bf 61}, 364 (2000).

\bibitem{stringari}
S. Stringari, Phys. Rev. B {\bf 49}, 6710 (1994).

\bibitem{wells} 
B. O. Wells {\it et al.}, Phys. Rev. Lett. {\bf 74}, 964 (1995); 
C. Kim {\it et al.}, {\it ibid.}, {\bf 80}, 4245 (1998).


\end{references}
\end{document}